\documentclass{pasj00}

\begin{document}
\SetRunningHead{Ueta et al.}{AKARI/FIS Mapping of $\alpha$ Ori}
\Received{2008/05/22}
\Accepted{2008/08/16}

\title{AKARI/FIS Mapping of the ISM-Wind Bow Shock around $\alpha$
Ori\thanks{Based on observations with AKARI, a JAXA project with
the participation of ESA.}} 

\author{%
Toshiya   \textsc{Ueta},\altaffilmark{1}
Hideyuki  \textsc{Izumiura},\altaffilmark{2}
Issei     \textsc{Yamamura},\altaffilmark{3}
Yoshikazu \textsc{Nakada},\altaffilmark{4,5}\\
Mikako    \textsc{Matsuura},\altaffilmark{6,\S}
Yoshifusa \textsc{Ita},\altaffilmark{3,6}
Toshihiko \textsc{Tanab\'{e}},\altaffilmark{4}
Hinako    \textsc{Fukushi},\altaffilmark{4}\\
Noriyuki  \textsc{Matsunaga}\altaffilmark{7,\S}
and
Hiroyuki  \textsc{Mito}\altaffilmark{5}}
\altaffiltext{1}{Department of Physics and Astronomy, University of Denver, 2112 E.\
Wesley Ave., Denver, CO 80208, USA}
\email{tueta@du.edu}
\altaffiltext{2}{Okayama Astrophysical Observatory, National
Astronomical Observatory, Kamogata, Asakuchi, Okayama 719-0232, Japan}
\altaffiltext{3}{Institute of Space and Aeronautical Science, Japan Aerospace
Exploration Agency, 3-1-1 Yoshinodai, Sagamihara, Kanagawa 229-8510, Japan}
\altaffiltext{4}{Institute of Astronomy, School of Science, 
University of Tokyo, 2-21-1 Osawa, Mitaka, Tokyo 181-0015, Japan} 
\altaffiltext{5}{Kiso Observatory, Institute of Astronomy, School of
Science, University of Tokyo, Mitake, Kiso, Nagano 397-0101, Japan} 
\altaffiltext{6}{National Astronomical Observatory of Japan, 2-21-1
Osawa, Mitaka, Tokyo 181-8588, Japan} 
\altaffiltext{7}{Department of Astronomy, Kyoto University,
Kitashirakawa Oiwake-cho, Sakyo-ku, Kyoto, Kyoto 606-8502, Japan}
\altaffiltext{\S}{JSPS Research Fellow}


\maketitle

\begin{abstract}
We present $10\arcmin \times 50\arcmin$ scan maps around an M supergiant
 $\alpha$ Ori in at 65, 90, 140 and 160$\mu$m
 obtained with the {\sl AKARI Infrared Astronomy Satellite}.
Higher spatial resolution data with the exact analytic solution permit
 us to fit the de-projected shape of the stellar wind bow shock around
 $\alpha$ Ori to have the stand-off distance of $4\farcm8$, position
 angle of $55^{\circ}$ and inclination angle of $56^{\circ}$. 
The shape of the bow shock suggests that the peculiar velocity of
 $\alpha$ Ori with respect to the local medium is $v_* = 40~n_{\rm
 H}^{-1/2}$, where $n_{\rm H}$ is the hydrogen nucleus density at $\alpha$ Ori.
We find that the local medium is of $n_{\rm H} = 1.5$ to 1.9 cm$^{-3}$
 and the velocity of the local flow is at 11 km s$^{-1}$ by using the
 most recent astrometric solutions for $\alpha$ Ori under the assumption
 that the local medium is moving away from the Orion OB 1 association.
{\sl AKARI} images may also reveal a vortex ring due to instabilities on
 the surface of the bow shock as demonstrated by numerical models.
This research exemplifies the potential of {\sl AKARI} All-Sky data as
 well as follow-up observations with {\sl Herschel Space Telescope} and 
{\sl Stratospheric Observatory for Infrared Astronomy} for
 this avenue of research in revealing the nature of interaction between
 the stellar wind and interstellar medium. 
\end{abstract}

\section{Introduction}

Alpha Orionis (Betelgeuse, HD 39801; hereafter $\alpha$ Ori) is a
supergiant of spectral type M2 Iab:\ and is a very bright far-infrared (far-IR)
source.
Some extended structure in the far-IR was recognized around
$\alpha$ Ori first by \citet{stencel88} in the All-Sky Survey data of
the {\sl Infrared Astronomical Satellite} ({\sl IRAS}\/).
Using an improved image reconstruction technique, \citet{nc97} produced
images of better quality for $\alpha$ Ori and identified the extended
structure around the star as an arc-shaped structure at the interface
between the stellar wind and the local interstellar medium (ISM).

Such structures can be formed around mass-losing stars at which the ram 
pressure of the ambient ISM balances with that of the stellar wind.
These stellar wind bow shock arcs had already been found around hotter,
more luminous OB stars in the {\sl IRAS} All-Sky maps \citep{vbm88}. 
Most recently, \citet{ueta06} discovered a stellar wind bow shock arc
around R Hya, an asymptotic giant branch star (an evolved star of lower
mass and lower rate of mass loss), using the {\sl Spitzer Space Telescope}.

Far-IR emission from these arcs is probably due mainly to thermal 
emission of cold dust components of the stellar wind bow shock whose
temperature peaks at far-IR.
There may be contribution from low-excitation atomic lines such as 
[O\emissiontype{I}] 63 $\mu$m, [O\emissiontype{I}] 145 $\mu$m and
[C\emissiontype{II}] 
158 $\mu$m at these wavelengths. 
However, the exact emission mechanism of these far-IR bow shocks
remains unclear, while an attempt to identify their spectroscopic
nature is currently on-going with the {\sl Spitzer Space Telescope}. 
Meanwhile, theoretical studies have been done both analytically
\citep{wilkin96, wilkin00} and numerically (e.g.\
\cite{ml91,dgani96,bk98,wareing07}) and the structure of the stellar
wind bow shocks has been understood reasonably well.

The {\sl AKARI Infrared Astronomy Satellite} ({\sl AKARI}\/;
\cite{murakami07}) makes the All-Sky Survey in the far-IR for the
first time in 25 years since {\sl IRAS} at much finer spatial
resolution.\footnote{%
The effective beam size of {\sl AKARI} ($0\farcm5$ to $0\farcm9$) is
much smaller than that of {\sl IRAS} ($2\farcm0$ to $5\farcm0$).}
The original {\sl IRAS} images of $\alpha$ Ori allowed only rough
identification of the shape of the extended far-IR emission
\citep{stencel88}.  
The enhanced {\sl IRAS} images of $\alpha$ Ori yielded just the mean
radius and width of the arc with a rough estimate for the position angle
of the apex \citep{nc97}.
In this paper, we characterize the structure of the far-IR stellar wind
bow shock around $\alpha$ Ori, using higher spatial-resolution {\sl
AKARI} images, and investigate the kinematics of the star, bow shock and
ISM in the vicinity of the star by adopting the analytic solution of the
bow shock \citep{wilkin96} and the most recent astrometric solutions for
the star \citep{harper08}. 

\section{Observations and Data Reduction}

We observed $\alpha$ Ori in the four bands at 65, 90, 140 and $160\mu$m
using the Far-IR Surveyor (FIS; \cite{kawada07}) on-board the
AKARI satellite \citep{murakami07} on 2006 September 21 as part of the
MLHES (``Excavating Mass Loss History in Extended Dust Shells of Evolved
Stars'') Mission Program (PI: I.\ Yamamura).
Observations were made with the FIS01 (compact source photometry) scan 
mode, in which two strips of forward and backward scans were done with 
a 70$\arcsec$ spacing 
at the $15\arcsec$ s$^{-1}$ scan rate, resulting in the sky coverage of
$10\arcmin \times 50\arcmin$ centered at the target.  

The FIS Slow-Scan Toolkit
(\cite{verdugo07}\footnote{Available at http://www.ir.isas.jaxa.jp/AKARI/Observation/};
ver.\ 20070914) was used to reduce the data.
We found that the quality of the resulting map was improved when we used
a combination of the temporal median filter with the width of 200 s (or
longer), temporal boxcar filter with the width of 90 s, and sigma
clipping threshold of 1.5. 
For the reduction of data in the short wavelength bands (SW bands; 65
and $90\mu$m), the results were improved further when we performed
flat-fielding using the local ``blank'' sky data.
Furthermore, we applied a custom reduction process to remove
pixel-dependent response variation by subtracting the baseline ``sky
value'' from each pixel determined by a linear least-squares fit to the 
off-source pixel values in the time-series data. 

The resulting maps are in 15 and $30\arcsec$ pixel$^{-1}$
(default pixel scale) for the SW and LW (long wavelength; 140 and
$160\mu$m) bands, respectively.
The resulting 1 $\sigma$ sensitivities are 4.0, 4.1, 6.5 and 19.7 MJy
sr$^{-1}$ while achieving, on average, five, eight, 15, and 10 sky
coverages per pixel were at 65, 90, 140 and $160\mu$m, respectively.
The sky emission (the component removed during the reduction) is found
to be $32.3\pm 4.6$, $31.7\pm 3.5$, $56.1\pm 5.8$ and $81.4\pm 9.8$ MJy
sr$^{-1}$ at 65, 90, 140 and 
$160\mu$m, respectively.
Photometry was done following the latest calibration method to address
the effects due to slow transient response of the Ge:Ga detectors
\citep{shirahata08}. 
Image characteristics are summarized in Table \ref{chara}.

The measured sky emission values are consistent with the estimates
obtained with the {\sl Spitzer Planning Observations Tool} ({\sl SPOT}\/) 
based on the {\sl COBE/DIRBE} data \citep{reach00} in the wide bands at
90 and $140\micron$ (28.7 and 52.1 MJy sr $^{-1}$, 
respectively), while about $50\%$ larger in the narrow bands at 65 and
$160\micron$ (21.7 and 53.8 MJy sr $^{-1}$, respectively).
By design, the spatial scale of the measured sky emission corresponds at
most to the scan length ($50^{\prime}$), which is comparable to the
spatial resolution of the far-IR background data used in the {\sl SPOT}
background estimates ($40^{\prime}$ to $70^{\prime}$; \cite{reach00}).
Hence, good agreement between the measured and estimated sky emission
values in the wide bands is reasonable. 
The $50\%$ discrepancy in the narrow bands, on the other hand, may
indicate highly structured and variable [O\emissiontype{I}] line
emission at 63 and $146\micron$ in the background in the vicinity of
$\alpha$ Ori, an oxygen-rich supergiant suffering from heavy mass loss.  
With our data, however, we are unable to either prove or disprove this
possibility, unfortunately.
Follow-up far-IR spectroscopic observations are indeed necessary.

\begin{table*}
\begin{center}
\caption{\label{chara}Characteristics of the AKARI/FIS Map of $\alpha$ Ori}
\begin{tabular}{lccccccc}
\hline
  & $\lambda$ & $\Delta\lambda$ & Pix Scale & Coverage & 
 $\sigma_{\rm sky}$ &
$S_{\nu, {\rm sky}}$ &
$S_{\nu, {\rm sky}}^{\rm SPOT}$\footnotemark[$*$] \\
 Band &  [$\mu$m] & 
 [$\mu$m] &  [arcsec] & [pixel$^{-1}$] & [MJy sr$^{-1}$] & [MJy sr$^{-1}$] & [MJy sr$^{-1}$] \\
\hline
N60    & \phantom{1}65 & 22 & 15 & \phantom{1}5 & \phantom{1}4.0 &
 $32.3\pm 4.6$ & 21.7 \\
WIDE-S & \phantom{1}90 & 38 & 15 & \phantom{1}8 & \phantom{1}4.1 &
 $31.7\pm 3.5$ & 28.7 \\
WIDE-L & 140 &           52 & 30 & 15           & \phantom{1}6.5 &
 $56.1\pm 5.8$ & 52.1 \\
N160   & 160 &           34 & 30 & 10           & 19.7 & $81.4\pm 9.8$ &
 53.4 \\
\hline
\multicolumn{8}{@{}l@{}}{\hbox to 0pt{\parbox{180mm}{\footnotesize
     \footnotemark[$*$] Estimate by {\sl SPOT} \citep{reach00}.
     }\hss}}
\end{tabular}
\end{center}
\end{table*}

\section{Results}

\begin{figure}
  \begin{center}
    \FigureFile(160mm,160mm){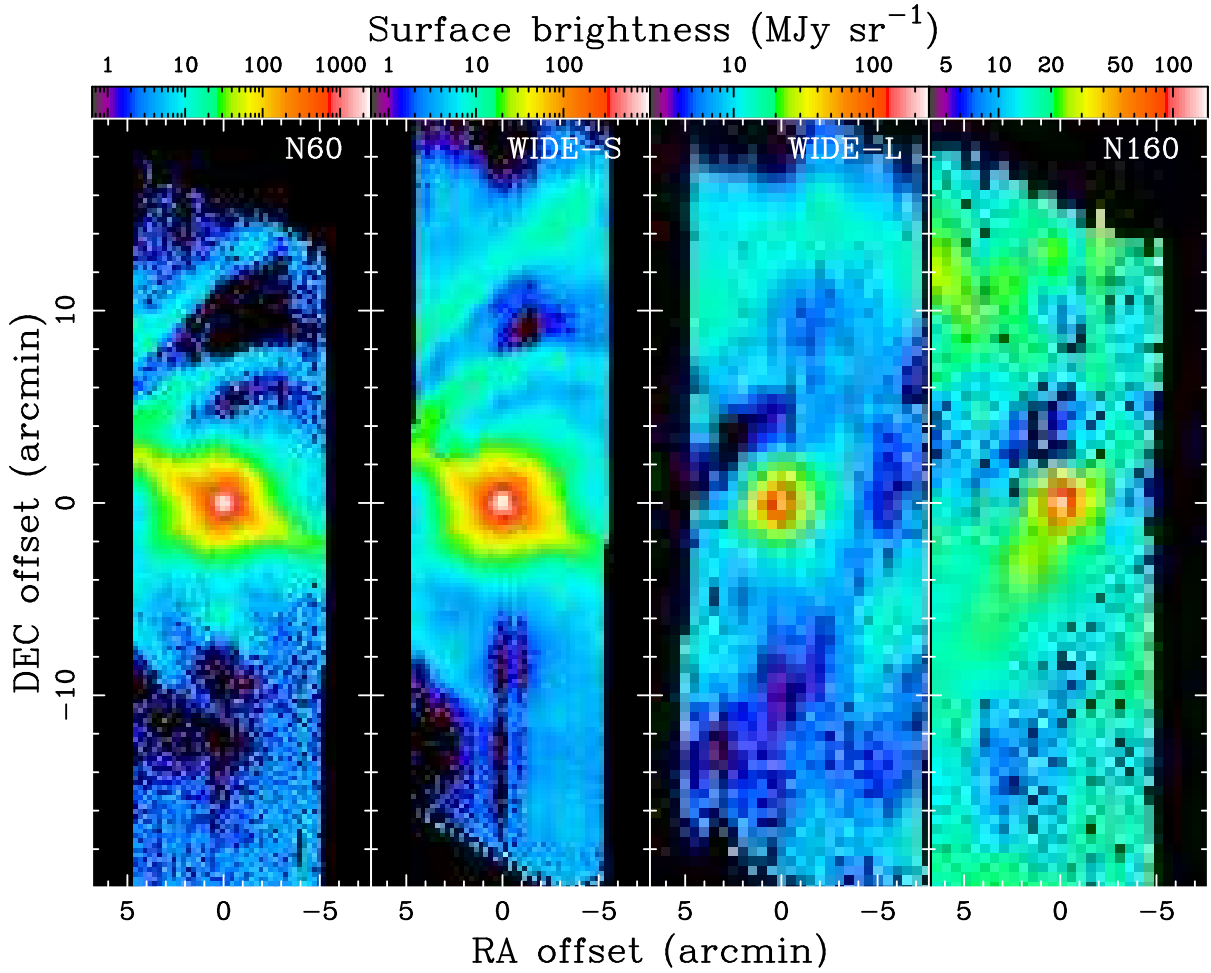}
  \end{center}
  \caption{\label{maps}AKARI/FIS false-color maps of $\alpha$ Ori in the
SW bands - N60 ($65\mu$m) and WIDE-S ($90\mu$m) at 15\arcsec
 pixel$^{-1}$ scale - and in the LW bands - WIDE-L ($140\mu$m) and N160
 ($160\mu$m) at 30\arcsec pixel$^{-1}$ scale - from left to right,
 respectively.  Background emission has been 
 subtracted by a combination of temporal filters during data
 reduction. RA and DEC offsets (with respect to the stellar peak) are
 given in arcminutes.  The wedges at the top indicate the log scale of
 surface brightness in MJy sr$^{-1}$.  North is up, and east to the left.}
\end{figure}

Figure \ref{maps} shows the background-subtracted false color AKARI/FIS
scan maps of $\alpha$ Ori at $65\mu$m (N60; {\it far left}), $90\mu$m
(WIDE-S; {\it second from left}), $140\mu$m (WIDE-L; {\it second from
right}) and $160\mu$m (N160; {\it far right}).
Surface brightness in MJy sr$^{-1}$ is indicated by the color scale that
is specified in the wedge above each scan map.
Due to the location of the detectors on the focal plane, the extent in
the in-scan direction is different in each band.
$\alpha$ Ori itself is clearly detected in all four bands.
However, the 65 and $90\mu$m band maps are affected by pixel
cross-talk induced by the bright star (linear extension of the star
oriented at 64$^{\circ}$ east of north) while the $160\mu$m band map is 
impacted by a ghost seen about $4\arcmin$ southeast of the star
\citep{kawada07}.
The extended central emission core is the topic of our forthcoming
paper and will not be discussed here. 

Measured fluxes of the star (aperture defined by where surface
brightness drops to the sky level) are 
$349 \pm 11$,
$151 \pm  4$,
$ 32 \pm  1$ and
$ 49 \pm  2$ Jy at 65, 90, 140 and $160\mu$m, respectively.
The flux values at 65 and $90\mu$m are consistent with the previous 
{\sl IRAS} measurements \citep{nc97}.
These values are obtained by following the standard method for
point-source photometry including aperture correction elucidated in the
Users Manual \citep{verdugo07}, except for the last step in which  
the effects due to slow transient response of the Ge:Ga detectors is
now addressed \citep{shirahata08}.
The AKARI Ge:Ga detectors are known to underestimate the flux by roughly
$60\%$ due to slow transient response, and the latest calibration allows
one to derive a correction factor via a power-law
function of the total flux (source plus background) based on the
calibration observations \citep{shirahata08}. 
Previously, the correction factors were assumed to be constants.

Besides the central star, the arc and bar are also clearly
detected in a much finer spatial scale than in the previous {\sl
IRAS} maps \citep{nc97}, even though the scan width was not wide enough
to capture these structures in their entirety.
Additional scans to cover the whole arc were scheduled on 2007 September
21.
Unfortunately, cryogenic liquid Helium boiled off on 2007 August 26 and
we were unable to obtain the additional data.
While detailed analysis of the structure of the arc has to wait until
the All-Sky data become available, we can still make use of the data at
hand. 

The arc is distinctly visible to the north and southeast of the star in
the 65 and $90\mu$m band maps and is marginally discernible in the $140\mu$m
band map.  
In the $160\mu$m band map, however, the arc is blended in with the background
cirrus and is not cleanly distinguishable.
The bar, on the other hand, can be seen in all four maps, while the
structure becomes progressively less well-defined at longer wavelengths.
In the SW band images, surface brightness of the brightest parts of the
arc (northeast of the star; 20 to 25 MJy sr$^{-1}$) is higher than that
of the bar ($\sim 10$ MJy sr$^{-1}$).
In the LW band images, however, surface brightness of the bar (15 to 20
MJy sr$^{-1}$) is higher than that of the arc (5 to 10 MJy sr $^{\-1}$).

Assuming that emission from the arc and bar detected in these maps is
mainly due to thermal emission of optically-thin concentration of dust
grains, we can estimate the dust temperature by fitting the observed
surface brightnesses at these bands to a Planck curve (i.e., $S_{\nu}
\propto \tau_{0}(\nu_0/\nu)^{\beta} B_{\nu} (T_{\rm dust})$, where
$\tau_{\nu} = \tau_{0}(\nu_0/\nu)^{\beta}$ is the optical depth at $\nu$
power-law scaled from $\tau_0$ at $\nu_0$ with an index $\beta$,
$B_{\nu}$ is the Planck function at $\nu$ and $T_{\rm dust}$ is the dust
temperature). 
 
Instead of treating $\beta$ as a free parameter, we varied it between 1
and 2 to see how $T_{\rm dust}$ and $\tau_0$ would behave.
We find $42 \pm 9$ K near the brightest parts to $11 \pm 1$ K near the
edge of the arc and $22 \pm 5$ K in the bar.
For the optical depth, we find values on the order of $10^{-5}$ to $5
\times 10^{-4}$ typically with a factor of $\sim 2$ uncertainty.
The optical depth in the arc is generally about an order of magnitude
lower than in the bar. 
Table \ref{arcbar} summarizes the observed characteristics of the arc,
bar and background sky.

\begin{table}
\begin{center}
\caption{\label{arcbar}Characteristics of the Arc and Bar Near $\alpha$ Ori}
\begin{tabular}{lcccc}
\hline
  & $\lambda$ & $S_{\nu}$ & $T_{\rm dust}$ & \\
 Object &  [$\mu$m] & 
 [MJy sr$^{-1}$] &  [K] & $\tau_{0}$ \\
\hline
Arc & \phantom{1}65 & $\phantom{1}5 ~{\rm to}~ 25$ & $11 ~{\rm to}~ 42$ & $4 \times
 10^{-5}$ \\
    & \phantom{1}90 & $\phantom{1}5 ~{\rm to}~ 20$ & \dots     & $2 \times 10^{-5}$\\
    & 140 &           $\phantom{1}5 ~{\rm to}~ 10$ & \dots     & $8 \times 10^{-6}$\\
    & 160 &           $10 ~{\rm to}~ 15$           & \dots     & $7 \times 10^{-6}$\\
Bar & \phantom{1}65 & $\sim 10$           & $17 ~{\rm to}~ 27$ & $5 \times
 10^{-4}$\\
    & \phantom{1}90 & $\sim 10$           & \dots     & $3 \times 10^{-4}$\\
    & 140 &           $\sim 15$           & \dots     & $1 \times 10^{-4}$\\
    & 160 &           $\sim 20$           & \dots     & $1 \times 10^{-4}$\\
Sky & \phantom{1}65 & $36.2$           & $25 ~{\rm to}~ 32$ & $5 \times
 10^{-4}$\\
    & \phantom{1}90 & $37.9$           & \dots     & $3 \times 10^{-4}$\\
    & 140 &           $59.7$           & \dots     & $1 \times 10^{-4}$\\
    & 160 &           $85.9$           & \dots     & $1 \times 10^{-4}$\\
\hline
\end{tabular}
\end{center}
\end{table}

\section{The Stellar Wind Bow Shock}

\subsection{Orientation of the Bow Shock Cone}

The far-IR arc around $\alpha$ Ori has been interpreted as the interface
between the interstellar medium (ISM) and the circumstellar envelope
developed by the stellar wind (\cite{stencel88,nc97}) at which the ram
pressure of the ambient ISM balances with that of the stellar wind, as
has been found around OB stars \citep{vbm88}.
The shape of such bow-shock arcs has been shown to follows the curve $z
= r^2/3 R_0$ (where $R_0$ is the stand-off distance between the star and
the apex of the bow and $r$ is the distance of the bow from the
symmetric $z$ axis of the bow) both numerically \citep{ml91} and
analytically \citep{wilkin96}. 

The shape of the arc around $\alpha$ Ori, however, appears to be much
more circular. 
If one traces the brightness peak of the arc to fit an ellipse, such an 
ellipse turns out to have the semi-major axis length of $9.6\arcmin$ and
the eccentricity of 0.02 centered at $4.0\arcmin$ off the star at the 
position angle of $56^{\circ}$ east of north. 
\citet{ml91} have already shown that the apparent shape of the bow
becomes more circular upon the consideration of the inclination angle. 
Since we have an exact analytic solution for stellar wind bow shocks
developed by \citet{wilkin96},
namely, 
\begin{equation}
 R(\theta)=R_0 \csc\theta\sqrt{3(1-\theta\cot\theta)}
\end{equation}
where $\theta$ is the polar angle from the apex of the bow, and the
observed shape of the arc is 
simply a conic section of an axisymmetric bow shock cone, we can fit
the brightness peak of the arc to this solution.

For a given set of the stand-off distance and inclination and position
angles of the bow (these angles define the 3-D orientation of the bow),
one can predict how the bow appears in the plane of the sky assuming that
(i) far-IR emission is mainly due to thermal dust emission,
(ii) the shell is optically thin to far-IR light and
(iii) the column density of dust is the highest where the bow
intersects with the plane of the sky.  
Dust radiative transfer calculations in the circumstellar shells have
shown that the above assumption (iii) is generally valid as long as the
assumption (ii) is valid (e.g.\ \cite{ueta01,ueta03}). 

Then, one can quantify the difference of the prediction from the data 
by computing the inverse of the sum of the squares of the differences
between the distance from the star to the arc peak and that from the
star to the predicted positions of the arc.
Since this quantity represents the ``correlation'' between the data and
prediction, it tends to be small if the prediction differs from what the
data suggest.
Thus, one can find the best-fit parameter set by locating the point in
the parameter space at which this quantity becomes the largest.

Through this method, we find 
the stand-off distance (de-projected) to be $4\farcm8 \pm 0\farcm1$, 
the position angle to be $55^{\circ} \pm 2^{\circ}$
and the inclination angle to be $56^{\circ} \pm 4^{\circ}$.
The uncertainties stem from those in determining the brightness peak of   
the arc by fitting the Gaussian to the surface brightness profile of the
$90\mu$m map, in which the arc is the most well-defined.
Figure \ref{arc} shows the position of the apex of the best-fit bow
shock cone projected in the plane of the sky.

The position angle is consistent with what has been concluded from
lower resolution {\sl IRAS} maps ($60^{\circ}\pm10^{\circ}$;
\cite{nc97})
and is now more accurately determined with higher resolution {\sl AKARI}
maps.
Adopting the heliocentric radial velocity of $v_{\rm rad} = +20.7\pm0.4$
km s$^{-1}$ (the mean of the measurements made by \citet{jones28} and
\citet{sanford33} as used by \citet{harper08}), we see that the bow
shock cone is oriented {\sl into} the plane of the sky, unlike the
previous studies assumed to be close to edge-on (e.g.\
\cite{nc97,harper08}).  
This is the first observational determination of the inclination angle of
a stellar wind bow shock, owing to higher resolution {\sl AKARI}
maps.

\subsection{Space Velocity of $\alpha$ Ori}

The stand-off distance is determined by the ram pressure balance of the
wind and ambient ISM. 
Therefore, starting from the pressure balance equation $\rho_{\rm w} v_{\rm
w}^2 = \rho v_{\rm *}^2$, one can determine the stand-off distance for
a star that loses mass via an isotropic stellar wind of velocity $v_{\rm
w}$ and mass-loss rate $\dot{M}$ while traveling 
through the ISM of a uniform density $\rho$ at velocity of $v_*$, 
the stand-off distance is
\begin{equation}
 R_0 = \sqrt{\frac{\dot{M}v_{\rm w}}{4\pi\rho v_*^2}}
\end{equation}
\citep{wilkin96}.
For $\alpha$ Ori, we have fairly well-established estimates for $\dot{M}
= 3.1 \pm 1.3 \times 10^{-6}$ M$_{\odot}$ yr$^{-1}$ \citep{harper01} and
$v_{\rm w} = 17 \pm 1$ km s$^{-1}$ \citep{bernat79}.
Also, \citet{harper08} have derived the new distance of $197 \pm 45$ pc
to $\alpha$ Ori based on astrometric solutions obtained by combining {\sl
Hipparcos} data with multi-epoch, multi-wavelengths VLA radio positions.
Using these values, the de-projected stand-off distance $R_0$ is $8.5
\pm 1.9 \times 10^{17}$ cm.
By keeping the interstellar hydrogen nucleus density at $\alpha$ Ori,
$n_{\rm H}$, as a free parameter, the peculiar velocity of the star {\sl
with respect to the ISM in the vicinity of $\alpha$ Ori} is 
\begin{equation}
 v_* = \sqrt{\frac{\dot{M}v_{\rm w}}{4\pi \mu_{\rm H} m_{\rm H} n_{\rm
  H} R_0^2}}
= (40 \pm 9)~n_{\rm H}^{-1/2} ~(\mathrm{km}~\mathrm{s}^{-1})\label{vpec}
\end{equation}
where $\mu_{\rm H}$ is the mean nucleus number per hydrogen nucleus for
local medium ($\sim 1.4$) and $m_{\rm H}$ is the mass of hydrogen
nucleus. 

Given the orientation of the bow shock cone, the peculiar velocity of
the star {\sl with respect to the ISM in the vicinity $\alpha$ Ori} can
be decomposed into each of the equatorial space-velocity components (the
radial direction, the direction in right ascension corrected for
declination and the direction in declination) as
\begin{equation}
 \left[
 \begin{array}{c}
   v_{\rho} \\
   v_{\alpha} \\ 
   v_{\delta} 
 \end{array}
 \right]^{\alpha~\mathrm{Ori}}_{\alpha~\mathrm{Ori~ISM}} = 
 \left[
 \begin{array}{c}
   33 \pm 8 \\
   18 \pm 5 \\
   13 \pm 3 
 \end{array}
 \right]~n_{\rm H}^{-1/2}~(\mathrm{km}~\mathrm{s}^{-1})
\end{equation}
These values can be converted to the Galactic space-velocity components
($U$, $V$ and $W$ where they are positive in the directions of the
Galactic center, Galactic rotation and the North Galactic Pole,
respectively) through 
a spherical trigonometric transformation (e.g.\ \cite{js87}) as
\begin{equation}
 \left[
 \begin{array}{c}
   U \\
   V \\
   W 
 \end{array}
 \right]^{\alpha~\mathrm{Ori}}_{\alpha~\mathrm{Ori~ISM}} = 
 \left[
 \begin{array}{c}
  -34 \pm 8 \\
  -10 \pm 4 \\
  +17 \pm 5 
 \end{array}
 \right]~n_{\rm H}^{-1/2}~(\mathrm{km}~\mathrm{s}^{-1})
\end{equation}
Since these values are based purely on the orientation of the bow shock
cone, they represent the Galactic space-velocity components {\sl of
$\alpha$ Ori with respect to the ISM in the vicinity of $\alpha$
Ori}.
Here, the superscript to the [$U$,$V$,$W$] vector indicates what
velocity it refers to while the subscript refers to with respect to what
the velocity is defined.

\subsection{ISM Flow in the Vicinity of $\alpha$ Ori}

The new astrometric solutions by \citet{harper08} yield the {\sl
heliocentric} Galactic space-velocity components {\sl of $\alpha$ Ori} as
\begin{equation}
 \left[
 \begin{array}{c}
   U \\
   V \\
   W 
 \end{array}
 \right]^{\alpha~\mathrm{Ori}}_{\odot} = 
 \left[
 \begin{array}{c}
  -22 \pm 1 \\
  -12 \pm 3 \\
  +21 \pm 4 
 \end{array}
 \right]~(\mathrm{km}~\mathrm{s}^{-1})
\end{equation}
Thus, by combining the above two sets of values we can compute the
{\sl heliocentric} Galactic space-velocity components {\sl of the ISM 
in the vicinity of $\alpha$ Ori} as 
\begin{equation}
 \left[
 \begin{array}{c}
   U \\
   V \\
   W 
 \end{array}
 \right]^{\alpha~\mathrm{Ori~ISM}}_{\odot} = 
 \left[
 \begin{array}{c}
  -22 \pm 1 \\
  -12 \pm 3 \\
  +21 \pm 4 
 \end{array}
 \right]
-
 \left[
 \begin{array}{c}
  -34 \pm 8 \\
  -10 \pm 4 \\
  +17 \pm 5 
 \end{array}
 \right]~n_{\rm H}^{-1/2}~(\mathrm{km}~\mathrm{s}^{-1})\label{flow}
\end{equation}
Hence, the stellar wind bow shock around $\alpha$ Ori is a consequence
of a mass-losing M supergiant moving in the ISM that flows in the 
direction specified by the above space-velocity vector.

This ISM flow around $\alpha$ Ori must originate from somewhere in the
vicinity of $\alpha$ Ori.  
The most probable source of this ISM flow in the vicinity of $\alpha$ Ori
is undoubtedly the Orion OB1 association (\cite{wh77}).
Assuming that all four sub-associations (OB1a through OB1d) contribute
equally to generate the ISM flow in the vicinity of $\alpha$ Ori, one
can define the flow vector toward $\alpha$ Ori at $[X, Y, Z] = [-182, -66,
-31$pc] emanating from the mean position of the associations at $[X, Y,
Z] = [-350,-170,-133$pc]. 

Then, one can search for values of $n_{\rm H}$ that would align the
ISM flow vector $[U,V,W]^{\alpha~\mathrm{Ori~ISM}}_{\odot}$ with this
flow from the Orion OB1 association 
within the quoted uncertainties in Eq.\ \ref{flow}.
Possible values of $n_{\rm H}$ turns out to be 1.5 to
1.9 cm$^{-3}$.  
These values are higher than $n_{\rm H} \simeq 0.3$cm$^{-3}$
estimated for material in front of the Orion OB association \citep{frisch90}. 
In the direction of $\alpha$ Ori, however, the H\emissiontype{I} column
density is estimated to be $0.27 \times 10^{22}$ cm$^{-2}$
\citep{kalberla05}, implying $<n_{\rm H}> = 4.4$ cm$^{-3}$ for the
distance of 197 pc.
Thus, our estimate of $n_{\rm H}$ may not be too large.

The {\sl heliocentric} Galactic space-velocity components {\sl of the ISM
flow in the vicinity of $\alpha$ Ori} can be converted back to the
equatorial 
space-velocity components through an inverse spherical trigonometric
transformation as
\begin{equation}
 \left[
 \begin{array}{c}
   v_{\rho} \\
   v_{\alpha} \\ 
   v_{\delta} 
 \end{array}
 \right]_{\odot}^{\alpha~\mathrm{Ori~ISM}} = 
 \left[
 \begin{array}{c}
 21 - 33~n_{\rm H}^{-1/2} \\
 23 - 18~n_{\rm H}^{-1/2} \\
 \phantom{1}9 - 13~n_{\rm H}^{-1/2} \\
 \end{array}
 \right]~(\mathrm{km}~\mathrm{s}^{-1})
\end{equation}
and for $n_{\rm H} = 1.5$ to 1.9 cm$^{-3}$
\begin{equation}
 \left[
 \begin{array}{c}
   v_{\rho} \\
   v_{\alpha} \\ 
   v_{\delta} 
 \end{array}
 \right]_{\odot}^{\alpha~\mathrm{Ori~ISM}} = 
 \left[
 \begin{array}{c}
-6 ~{\rm to}~ -3 \\
\phantom{-}9 ~{\rm to}~ 10 \\
-1 ~{\rm to}~ \phantom{1}0 \\
 \end{array}
 \right]~(\mathrm{km}~\mathrm{s}^{-1})
\end{equation}
Thus, the ISM around $\alpha$ Ori flows at about 11 km s$^{-1}$ into the 
position angle of $\sim 95^{\circ}$ out of the plane of the sky
(toward us).

Since the stellar wind bow shock is a consequence of the peculiar motion
of the star {\sl and} the ISM flow at the star, the apparent orientation
of the bow shock would not necessarily yield information on both.
When the direction of the proper motion is aligned with the
direction of the apex of the bow, the ISM at the star is stationary with  
respect to the Sun or flows against the peculiar motion of the star.
On the other hand, when the direction of the proper motion is {\sl not}
aligned with the direction of the apex of the bow, the ISM at the star
probably flows obliquely with respect to the peculiar motion of the
star. 
Therefore, direct comparison of the orientation of the bow with respect
to the direction of the peculiar motion of the star would provide a
reasonable diagnostic for the presence and direction of the ISM
flow around the star.

\subsection{Substructure of the Bow Shock}

For the derived values of $n_{\rm H} = 1.5$ to 1.9 cm$^{-3}$, the
corresponding peculiar velocity of the star {\sl with respect to the 
ISM in the vicinity of $\alpha$ Ori} is 33 to 29 km s$^{-1}$ (Eq.\ref{vpec}). 
This means that the ratio of the peculiar velocity of the star to the
wind velocity ($v_{\rm *}/v_{\rm w}$) is $1.7$ to 1.9.
Thus, the bow shock may be prone to instability (\cite{dgani96,bk98})
and may even develop vortices (e.g.\ \cite{wareing07}).

\begin{figure}
  \begin{center}
    \FigureFile(80mm,80mm){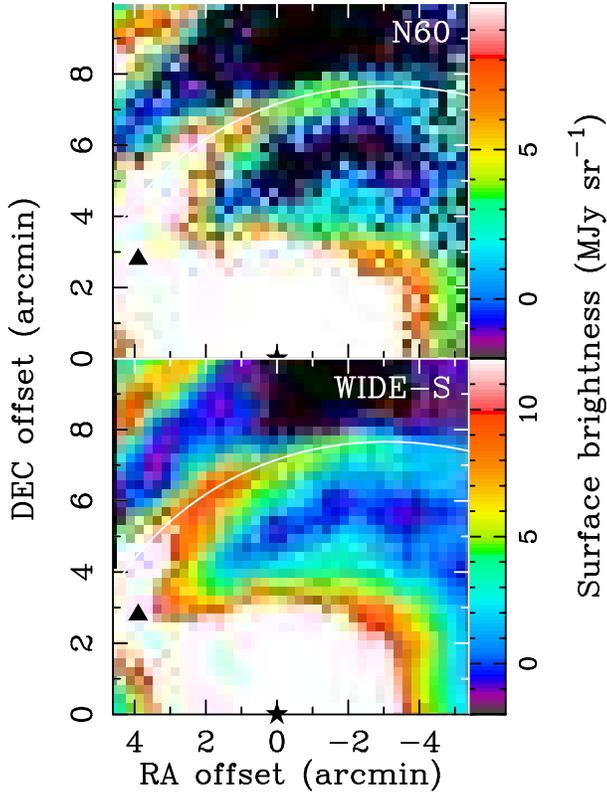}
  \end{center}
  \caption{\label{arc}Close-up of the northern arc structure in the
 WIDE-S ($90\mu$m) map.   
 RA and DEC offsets (with respect to the stellar peak) are
 given in arcminutes.  The wedges at the top indicate the log scale of
 surface brightness in MJy sr$^{-1}$.  North is up, and east to the
 left.
 The white line is the Wilkin (1996) curve that
 delineates the analytic solution at the 55$^{\circ}$ position angle at
 the $47^{\circ}$ inclination angle.
 The black star and triangle indicate the position of $\alpha$ Ori and
 of the apex of the best-fit bow shock cone projected in the plane of
 the sky, respectively.} 
\end{figure}

Figure \ref{arc} is a close-up of the northern part of the arc overlaid
with a line that delineates the \citet{wilkin96} analytic solution at
the 55$^{\circ}$ position angle at the $47^{\circ}$ inclination angle. 
In general the Wilkin curve predicts where the surface brightness peaks
along the arc extremely well.
However, as one follows the arc structure from the apex to the
downstream direction, there is a discontinuity of surface brightness
along the arc at the position angle $\sim 0^{\circ}$ and a local
enhancement of surface brightness at around (0\arcmin, 7\arcmin) from
the star.
The position of this local brightness enhancement is somewhat interior
to the Wilkin curve unlike other parts of the arc.

This drop of surface brightness accompanied by a local enhancement off
the Wilkin curve may be due to vortex shedding caused by instabilities
in the bow surface (\cite{bk98,wareing07}).
In an isothermal bow shock, \citet{bk98} found that instabilities would
manifest themselves as wiggles in the bow on the length scale of the
stand-off distance. 
Figure \ref{arc} shows that reduction and enhancement of surface
brightness occurs about the stand-off distance away from the apex.
Qualitatively, the observed structure of the bow shock resembles to that 
of the numerical models showing vortex shedding and the development of
vortex rings.
Therefore, more detailed numerical investigations into the development
of instabilities on the surface of the bow shock appears worthwhile.

\subsection{The Bar ahead of the Bow Shock Cone}

The existence of a linear bar structure ahead of $\alpha$ Ori is
intriguing. 
At the moment there is no evidence that indicates the bar being a
(by)product of the interaction between the ISM and stellar wind from
$\alpha$ Ori or otherwise:
the origin of the bar remains unclear.
Similarly, there is no evidence that suggests the bar is co-spatial with
$\alpha$ Ori.
However, our data (Table \ref{arcbar}) show higher optical depths in the bar
than in the arc and in the background.
Moreover, the bar and background seem to have roughly the same optical
depth values.

Thus, the bar appears to be more like the background cirrus than the
arc, and hence, the bar is probably not caused by the interaction
between the stellar wind and ISM.
Also, assuming that these structures are made up with similar matter and
have similar density, the differences in the optical depth suggests that
the bar is extended along the line of sight like a sheet.
If the bar represents a local concentration of matter co-spaced
with $\alpha$ Ori, the motion of the star and angular separation between
the star and bar imply a collision between the two in about 3000 yr.
Even if this is the case, it is still not clear whether the bar is
caused by this ISM flow in this region or is caused by other external
means (such as $\lambda$ Ori).

\section{Summary}

AKARI/FIS scan maps around $\alpha$ Ori ($10\arcmin \times 50\arcmin$)
at 65, 90, 140 and $160\mu$m are presented.
These images show the extended emission core and most of the
circumstellar arc plus the northern bar structure at much higher spatial
scale than in the previously obtained {\sl IRAS} maps.
Spatial resolution of the data is good enough to define the structure of
the arc to be fit with the exact analytic solution for stellar wind bow
shocks, while the scan did not cover the arc in its entirety and there
are some anomalies by pixel cross-talk in the SW band and by ghosting in
the $160\mu$m band due to the bright central star.

Rather circular appearance of the arc (eccentricity 0.02) suggests that 
the stellar wind bow shock cone of $4\farcm8$ de-projected stand-off
distance is inclined at $56^{\circ}$ with respect to the plane of the
sky and oriented at $55^{\circ}$ position angle (east of north).
Adopting the distance of 197 pc, rate of mass loss at $3.1 \times
10^{-6}$ M$_{\odot}$ yr$^{-1}$, and wind velocity of $17$km s$^{-1}$,
the peculiar velocity of the star with respect to the ISM at $\alpha$
Ori is found to be $v_{*} = 40~n_{\rm H}^{-1/2}$km s$^{-1}$.

By comparing the Galactic space-velocity components of the peculiar
velocity of the star (based on the orientation of the bow shock cone)
and that of the apparent motion of the star (based on the astrometric
solutions), we derived the space-velocity components of the ISM
flow around the star. 
Assuming that the ISM flow in the vicinity of $\alpha$ Ori emanates
from the Orion OB 1 association, we find the particle
number density per hydrogen nucleus ($n_{\rm H}$) in the ISM around
$\alpha$ Ori to be 1.5 to 1.9 cm$^{-3}$, which translates to a 11 km
s$^{-1}$ flow. 

Owing to higher spatial resolution of the data, we may be witnessing the
development of a vortex ring along the surface of the bow due to
instabilities. 
This research demonstrates that far-IR images of stellar wind bow shocks
are excellent diagnostic tools to investigate the kinematics of the bow
and the characteristics of the ISM in the vicinity of the star.
These stellar wind bow shocks have been found not only around M
supergiants but also around OB stars \citep{vbm88} and an AGB star
\citep{ueta06}. 
Therefore, this avenue of research is going to flourish with the coming 
of the AKARI All-Sky Survey \citep{murakami07} followed by new
opportunities with {\sl Herschel Space Telescope} and {\sl Stratospheric
Observatory for Infrared Astronomy}, because there will 
be a wealth of new far-IR data on the stellar wind bow shocks around
variety of mass-losing stars. 





\bigskip
This research has made use of the SIMBAD database,
operated at CDS, Strasbourg, France.
Ueta is grateful to Drs.\ R.\ E.\ Stencel and G.\ M.\ Harper for
illuminating discussions. 
Ueta also acknowledges contribution by a summer student, A.\ Karska, in  
developing some of the custom data reduction algorithms.


\clearpage

\clearpage

\end{document}